\documentclass{aastex}
\slugcomment{}

\def\deg{$^{\circ}$}
\def\kms{km\thinspace s$^{-1}$}
\def\cm2{cm$^{-2}$}
\font \sc = cmr10

\def\sun{_\odot}
 
\shortauthors{Walter et al.}
\shorttitle{ISM and Star Formation in NGC~4214}

\begin{document}

\title{The Interaction between the ISM and Star Formation in the Dwarf 
Starburst Galaxy NGC~4214}

\author{Fabian Walter}
\affil{California Institute of Technology}
\authoraddr{California Institute of Technology, Astronomy Department 105-24, 
Pasadena, CA, 91125}

\author{Christopher L. Taylor}
\affil{Five College Radio Astronomy Observatory}
\authoraddr{University of Massachusetts, Five College Radio Astronomy 
Observatory, 619B Lederle GRT, Amherst, MA 01003 }

\author{Susanne H\"uttemeister}
\affil{Radioastronomisches Institut, Universit\"at Bonn}
\authoraddr{Universit\"at Bonn, Radioastronomisches Institut, Auf dem 
H\"ugel 71, 53121 Bonn, Germany }

\author{Nick Scoville}
\affil{California Institute of Technology}
\authoraddr{ California Institute of Technology, Astronomy Department 105-24,
Pasadena, CA, 91125}

\author{Vincent McIntyre}
\affil{Australia Telescope National Facility, CSIRO}
\authoraddr{PO Box 76, Epping, NSW 1710, Australia}

\vspace{.75in}

\begin{abstract}

We present the first interferometric study of the molecular gas in the
metal--poor dwarf starburst galaxy NGC\,4214. Our map of the $^{12}$CO
(1--0) emission, obtained at the OVRO millimeter array, reveals an
unexpected structural wealth.  We detected three regions of molecular
emission in the north--west (NW), south--east (SE) and centre of
NGC\,4214 which are in very different and distinct evolutionary stages
(total molecular mass: $\sim 5.1 \times 10^6 M_{\odot}$).  These
differences are apparent most dramatically when the CO morphologies
are compared to optical ground based and HST imaging: massive star
formation has not started yet in the NW region; the well--known
starburst in the centre is the most evolved and star formation in the
SE complex started more recently.  We derive a star formation
efficiency of $\sim8\%$ for the SE complex. Using high--resolution VLA
observations of neutral hydrogen (\ion{H}{1}) and our CO data we
generated a total gas column density map for NGC\,4214 (\ion{H}{1} +
H$_2$). No clear correlation is seen between the peaks of \ion{H}{1},
CO and the sites of ongoing star formation.  This emphasizes the
irregular nature of dwarf galaxies.  The \ion{H}{1} and CO velocities
agree well, so do the H$\alpha$ velocities.  In total, we cataloged 14
molecular clumps in NGC\,4214. Our results from a virial mass analysis
are compatible with a Galactic CO-to-H$_2$ conversion factor for
NGC\,4214 (lower than what is usually found in metal--poor dwarf
galaxies).

\end{abstract}

\section{Introduction}

Previous studies have shown that molecular gas is hard to study in
dwarf galaxies, even in those who are known to be rich in neutral
atomic gas (\ion{H}{1}), such as dwarf irregulars (dIrrs) and blue compact
dwarfs (BCDs).  As carbon monoxide (CO) is the second most common
molecular species after molecular hydrogen (H$_2$), and the one most
easily observed in its cool, unshocked state, it is usually used as a
tracer of molecular gas.  Relatively few dwarf galaxies have been
detected in CO, though many have been observed (e.g., Taylor,
Kobulnicky \& Skillman 1998, Barone et al.\ 2000).  Dwarf galaxies are
set apart from from other \ion{H}{1}-rich galaxies by their low metallicity:
Recent work suggested that the CO-to-H$_2$ (`X') conversion
factor is dependent upon the metallicity of the gas, with the
conversion factor increasing as metallicity decreases (Arimoto, Sofue
\& Tsojimoto 1996, Wilson 1995, Verter \& Hodge 1995, Rubio et al.\
1991).  Maloney \& Black (1988) argued on theoretical grounds that at
low metal abundances, the column density of CO will be lower, and
there will be less self shielding from dissociating radiation.  This
will cause the size of the CO emitting region to shrink, while leaving
the H$_2$ unaffected.  Norman \& Spaans (1997) predict that
metallicities starting somewhere between 0.03 to 0.1 of the solar
value are necessary before CO will be detectable.  Observationally,
Taylor et al.\ (1998) have found no dwarf galaxies with detectable CO
emission at metallicities below $\sim$ 0.15 solar. However, at present
there are too few dwarfs with both metallicity measurements and CO
observations to determine reliably if there is indeed a cutoff, and
where exactly it might take place.

For the few dwarfs which have been detected in CO, following up those
single dish detections with high spatial resolution interferometer
data has proven enlightening.  In some cases, the high resolution of
interferometers allow the direct estimate of the total mass of
resolved giant molecular clouds in nearby dwarf galaxies independent
of the CO intensity, and from this a first determination of the
CO--to--H$_2$ conversion factor (e.g., Wilson 1995).  A list of dwarf
galaxies which have been observed with interferometers to date is
compiled in Tab.~1. E.g., Taylor et al.\ (1999) used the Plateau de
Bure interferometer to observe CO (1--0) and (2--1) in the
post-starburst dwarf galaxy NGC~1569.  They found that the conversion
factor was three times higher than in two other dwarf galaxies of the
same metallicity (NGC\,6822 and IC\,10), suggesting that metallicity
is not the only influence upon the conversion factor.  Motivated by
this and other previous studies, we have observed another starburst
dwarf galaxy, NGC~4214, at Owens Valley.

\tablenum{1}
\placetable{table1}
\label{table1}

NGC~4214 is a nearby dwarf galaxy currently experiencing high levels
of massive star formation.  Table~2 summarizes some of the properties
of NGC~4214. Combining optical, NIR and UV data, Huchra et al.\ (1983)
concluded it went through a burst of star formation a few
$\times~10^7$ yr ago. NGC~4214 has two main regions of star formation
-- a large complex of \ion{H}{2} regions displaying a shell morphology in the
galaxy center, and a smaller southern complex formed mainly of compact
knots (MacKenty et al.\ 2000). From age determinations of the \ion{H}{2}
regions and their spatial distribution, they argue that some of the
younger star formation has been triggered by the interaction of the
older \ion{H}{2} regions with the surrounding ISM.  At a distance of 4.1 Mpc
(Leitherer et al.\ 1996), it is one of the closest starbursting dwarf
galaxies, and thus an excellent target for a high resolution study of
the molecular ISM.  Several groups have made single dish CO
observations of NGC~4214 (Tacconi \& Young 1985, Thronson et al.\
1988, Ohta et al.\ 1993, Becker et al.\ 1995, Taylor et al.\ 1998).
While the existence of CO emission away from the center of NGC~4214
can be inferred from the single dish data of Ohta et al.\ (who,
however, fail to detect the central emission peak) and Taylor et al.,
no analysis of the distribution of the molecular emission and its
relation to the major star forming regions of the galaxy has been
presented to date. Thronson et al.\ (1988) have also measured the dust
mass in a region about $4'$ in size using 160 $\mu$m data.  The \ion{H}{1} has
been studied by Allsopp (1979) and McIntyre (1998) who find a large
extended \ion{H}{1} disk and a complex velocity field in the center. Based on
the disturbed inner velocity field, Allsopp (1979) argued that
NGC~4214 is interacting with NGC~4190, a dIrr with a projected
distance of 35 kpc.

\tablenum{2}
\placetable{table2}
\label{table2}

In Sec.~2 we describe our CO observations obtained with OVRO. In
Sec.~3 we discuss the properties of the CO emitting regions detected
by the interferometer, compare them to high--resolution VLA \ion{H}{1} imaging
as well as optical ground--based and HST imaging of NGC\,4214. In
Sec.~4 we will discuss the conversion factor, the star formation
efficiency, implications on the star formation threshold as derived
from our CO and \ion{H}{1} data as well as the interplay of the star formation
regions on the surrounding ISM in NGC\,4214. We summarize our results
in Sec.~5.

\section{OVRO Observations}

\subsection{Observations and Data Reduction}

We observed NGC\,4214 in the CO(1--0) transition using the OVRO
millimeter interferometer in mosaicing mode in C and L
configuration. In total, 50 hours were spent on source. The
observational details are listed in Table~3. Data were recorded using
two correlator setups resulting in a velocity resolution of 5 and 1.3
\kms\ (after online Hanning smoothing) with a total bandwidth of 320
and 80 km\,s$^{-1}$, respectively. The absolute flux calibration was
determined by observing 3C273, 3C345 and (in some observing periods) 
Neptune for approximately 20 minutes during each observing run. 
These calibrators and an additional noise source were used to derive 
the complex bandpass corrections. The quasar 1156+295, which is close
to NGC~4214 on the sky, was used as secondary amplitude and phase 
calibrator. During the program, its flux decreased from $\sim 4.0$\,Jy 
(Sep 1999) to $\sim 1.0$\,Jy (June 2000).

The data for each array were edited and calibrated separately with the
{\sc MMA} and the {\sc MIRIAD} packages. The {\it uv--}data were
inspected and bad data points due to either interference or shadowing
between telescopes were removed, after which the data were
calibrated. We Fourier transformed our C and L array observations
separately for each pointing to assess their quality; subsequently we
combined all data to form a single dataset which was used for
mapping. The mosaicing and mapping was performed using the task {\sc
MOSSDI} in {\sc MIRIAD}, i.e.\ a joint deconvolution of the entire
mosaic was done to obtain the final data cubes. All the results
presented in this paper are based on this combined dataset.

We calculated 2 data cubes, one with natural weighting, leading to a
resolution of $6.4'' \times 5.7''$, and another one with robust
weighting (setting the robust parameter to 0, Briggs 1995) with a
resolution of $4.8'' \times 4.2''$. The final rms noise in our
1.3\,\kms\ wide channels is 40 mJy\,beam$^{-1}$ (100\,mK) and 47
mJy\,beam$^{-1}$ (210\,mK) for the naturally and robustly weighted
cube, respectively. We optimized the cleaning process by defining 3
clean boxes around the prominent emission regions.  A summary of the
observations can be found in Tab.~3. Channel maps of the data cube are
presented as contours in Fig.~5.

\tablenum{3}
\placetable{table3}
\label{table3}

\subsection{Data Reduction}

We distinguished CO emission from the noise using the following
procedure: The cleaned 1.3 \kms\thinspace data cube was smoothed to a
spatial resolution of 9\arcsec~$\times$~9\arcsec, and a 2$\sigma$ (80
mJy beam$^{-1}$) blanking was applied to the output cube.  The
resulting cube was searched for CO emission by looking for surviving
features which were continuous in at least 3 consecutive channels.
This gave a cube containing only the CO emission at a resolution of
9\arcsec.  This cube was then applied as a mask to the original
resolution naturally weighted cube to yield the final blanked data
cube.  The mask was also applied to the robustly weighted cube, and
moment maps were made from the data of both weightings (Fig.~1).

The locations and sizes (FWHM) of the OVRO primary beams comprising
our mosaic are shown in Fig.~7.  While we cannot be certain to have
detected all the CO emission in NGC~4214, we have probably mapped all
the strong emission.  Our first set of observations included only the
center mosaic position, but in those data we still detected both the
northwestern and southeastern regions, even though they lay beyond the
40\% power radius of the primary beam, therefore if there were any
similar CO features just beyond the beam areas shown in Fig.~7, we
would have detected them.

It would be interesting to estimate the degree of missing flux due to
the lack of short spacings in our interferometric map. This might give
some indication of the amount and distribution of diffuse gas in and
between the complexes. Unfortunately, there are no single dish
observations of the CO distribution in NGC~4214 that are suitable for
such a determination.  The single dish beam is either very large
($50'' - 55''$) with a pointing center offset from the peak position
of any complex, thus partially containing several complexes (Taylor et
al.\ 1998, Thronson et al.\ 1988, Tacconi \& Young 1985), or the
observed region is not sufficiently extended and spatially disjunct
(Ohta et al.\ 1993), or the observations were done in a different
transition (Becker et al.\ 1995 for the central complex). 

\placefigure{fig1_rev}

\placefigure{fig2_rev}

\section{Results}

\subsection{CO Emitting Regions}

As seen in the moment map presented in Fig.~1, there are three regions
where we have detected CO emission in NGC~4214 -- one at the center of the
galaxy, another $\sim$ 650 pc to the southeast of the center, and the
third $\sim$ 760 pc to the northwest.  The southeastern region (SE) has the
highest peak in CO emission, and is also the most centrally concentrated.
The central region is extended in the east-west direction,
with low intensity CO emission that has a diffuse appearance except for 
the concentration at its western end. The northwestern region (NW) is 
intermediate between the central and the southeastern ones,  both in 
the degree of central concentration and the intensity of the emission.  

Each of the regions can be divided into discrete units (clumps) by
considering both spatial and velocity information given in the data
cube. To do that, we inspected the high--resolution robust--weighted
cubes by looking at `movies' as well as position--velocity (pv)
diagrams. The clump decomposition is difficult in the NW and SW
regions where many smaller units are blended. Fig.~2 shows a pv
diagram of the NW region as an example of how we defined the clumps
(the ellipses plotted in this figure are only used to label the
clumps). The properties of the identified CO emitting clumps in
NGC~4214 are compiled in Table~4. C1--C8 belong to the NW complex, C9
belongs to the central complex, where also some diffuse emission is
present (D1) and C10--C14 are the constituents of the SE emission
feature in NGC~4214.

For each clump, we derived a total intensity map, measured the
diameter defined by the half intensity line and the velocity width:
Tab.~4 lists the central coordinates (column 2 and 3), the central
velocity of the clump (column 4), the velocity width (FWHM, column 5),
the CO flux (column 6) as well as the diameter of the clump (FWHM,
column 7) along the cut deconvolved for our beamsize (average of the
orthogonal directions). In many cases, only an upper limit for the
cloud size could be determined (as given by the size of the robust
beam). We also attempted to estimate molecular masses by employing a
Galactic X--factor (column 8, see Sec.~\ref{xfact}) as well as
assuming virialization (column 9, but see the discussion in
Sec.~\ref{xfact}).

C1--C14 are likely not Giant Molecular Clouds (GMCs).  This can be
seen in Fig.~3, where we show the size-linewidth relation for GMCs in
M33 from Wilson \& Scoville (1990), along with GMCs from M33, M31
(Vogel et al.\ 1987, Wilson \& Rudolph 1993), the SMC (Rubio et al.\
1993), IC10 (Wilson 1995), NGC\,1569 (Taylor et al. 1999) and NGC~6822
(Wilson 1994), and our complexes.  The three complexes resolved by the
OVRO beam clearly occupy a different region in the plot than GMCs in
Local Group galaxies.  They have similar linewidths, but are larger in
size.  The unresolved clouds may be individual GMCs, but their
diameters are too small to measure with the synthesized beam of our
data.

The clumps we see fall in an area in the size-linewidth space which is
in between the Local Group Galaxies and CO complexes in M83 (Rand,
Lord \& Higdon 1999), also plotted in Fig.~3.  Rand et al. argue that
their objects are intermediate in size, linewidth, and molecular mass
between the largest GMCs known in the Milky Way, and Giant Molecular
Associations (GMAs) seen in nearby spiral galaxies (e.g. Rand \&
Kulkarni 1990).  Our resolved objects have smaller diameters compared
to the Rand et al.\ objects -- they are probably not GMAs. Also, GMAs
tend to have masses of a few $\times~10^7$ M$_\odot$, whereas for an
assumed Galactic CO-H$_2$ conversion factor, our clumps have masses
that are lower by almost two orders of magnitudes.

\placefigure{fig3_rev}

\tablenum{4}
\placetable{table4}
\label{table4}

\subsection{Comparison to the \ion{H}{1}}

Starbursting dwarf galaxies tend to be \ion{H}{1} rich (e.g. Thuan \& Martin
1981, Taylor et al.\ 1995), and NGC~4214 is no exception.  Fig.~4
shows the CO contours from Fig.~1 superposed on the total intensity
map of the \ion{H}{1} emission (corrected for primary beam attenuation).  The
high--resolution \ion{H}{1} data are combined B, C and D configuration
observations from the VLA, with a spatial resolution of
8$\arcsec~\times~8\arcsec$ (see also McIntyre 1998). The area we have
mapped in CO covers only a small region at the center of NGC~4214.
The peak column density of \ion{H}{1} in the map is $\sim$
3~$\times~10^{21}$\,\cm2, and the overall \ion{H}{1} distribution shows a
large degree of structure: \ion{H}{1} holes and shells, which are common in
dwarf galaxies (e.g. Puche et al.\ 1992, Walter \& Brinks 1999), and
some indication for spiral structure extending beyond the main optical
disk. There is even a hint of a bar in the \ion{H}{1}, running roughly
position angle 45\deg\ (N $\to$ W) through the center of the galaxy.
It is possible that both the weak spiral features and the bar are
consequences of an interaction with the nearby dwarf galaxy
NGC~4190. There are a few dwarf galaxies such as Holmberg\,II (Puche
at al. 1992) and DDO\,47 (Walter \& Brinks 2001) which show similar
spiral features in \ion{H}{1} only. The underlying process causing transient
spiral structure may thus be fairly common.

A global comparison between the \ion{H}{1} and CO emission reveals differences
between the various CO emitting regions.  The NW region of CO emission
falls on top of a local maximum in the \ion{H}{1} column density, while the SE
region, which is brighter and more centrally concentrated, corresponds
to a dip in the \ion{H}{1} column density near two local peaks. The central CO
region is extended enough to cover several local minima and maxima in
the \ion{H}{1}, though the shape of the \ion{H}{1} is elongated east-west, similar to
the CO emission. It should be noted here that the strongest molecular
emission is not necessarily found where the \ion{H}{1} column densities are
highest. This finding emphasizes the `irregular' nature of dIrrs.

\placefigure{fig4_rev}

The CO and \ion{H}{1} channel maps are compared in Fig.~5.  This
representation of the data emphasizes more strongly the location of
the SE CO emission in a small \ion{H}{1} dip, and the association of
the NW CO emission region with a local \ion{H}{1} maximum roughly coincident
in velocity.  There is no clear correlation in the channel maps between 
CO and \ion{H}{1} for the central CO feature.  Fig.~6 is a pv-diagram through
the three brightest CO peaks and compares the CO and \ion{H}{1} velocities and
intensities in and near these peaks. For the SE (offset:
--0.75$'$) and central peak the CO linewidths are much narrower than
the corresponding \ion{H}{1} linewidths.  In contrast, in the the NW peak
(at offset 0.7$'$) the CO linewidth is larger, probably due to substructure
in the complex, and approaches the velocity width of the \ion{H}{1} to within a 
factor of $\sim 1.5$.

\placefigure{fig5_rev}

\placefigure{fig6_rev}

\subsection{Comparison to Stellar Component and Regions of Star Formation}

Star formation occurs at densities where the gas is primarily
molecular, so it is of interest to compare the molecular gas tracer,
CO, to tracers of the stellar component.  Because the ISM is readily
ionized by the intense UV radiation from young stars in \ion{H}{2} regions,
H$\alpha$ is often used as an indicator of the presence of recent star
formation.  Fig.~7 compares the CO emission to an H$\alpha$ image
(courtesy C. Martin).  The left panel shows lacy, filamentary
structure outside the immediate vicinity of the star forming regions.
These are similar to those seen in other dwarf galaxies
(e.g. NGC~1569; Hunter et al.\ 1993, Devost et al.\ 1997, Martin
1998), and are presumably caused by the energetic by-products of star
formation, supernovae and stellar winds.

Two of the three CO complexes are associated directly with the star 
forming regions (Fig.~7, right).  The SE CO clump appears cospatial 
with a large star forming complex containing several \ion{H}{2} regions.  The 
peak of the CO emission here is almost directly on top of one of the \ion{H}{2}\
regions, and the shape of the CO complex with its slight extension in the
north-south direction is very similar to that of the \ion{H}{2}\ region.
The central CO emission is associated with the largest region of
H$\alpha$ emission.  Unlike the SE complex, the central CO complex 
is diffuse instead of centrally concentrated, and the peak of the 
CO emission here is not directly on top of the H$\alpha$ emission, 
but is shifted to the west of it.  Lower level CO emission is cospatial
with the H$\alpha$, however, tracing out the chain of brighter
\ion{H}{2}\ regions running east-west in this area.  In clear contrast, only 
very faint H$\alpha$ emission is associated with the NW CO complex. 

At this point it is interesting to compare the velocities of the
molecular gas with the ones of the ionized component (as seen in
H$\alpha$ spectra). Hartmann et al. (1986) presented the kinematics of
a few \ion{H}{2} regions within NGC\,4214 -- four of them correspond
to regions where we also detected CO emission (their regions: H25/29,
close to our NW CO complex; H36 and H48, towards the central complex;
and H55 in the SE complex). In all cases do the systemic CO and
H$\alpha$ velocities agree well (this is consistent with the results
by Maiz--Apellaniz et al. 1999). Only in H25/29 (the weakest of the
four H$\alpha$ regions) is H$\alpha$--emission also present at
velocities where no CO--emission is present (around v$_{\rm
hel}$=250--270\,\kms).

Fig.~8 shows the contours of the CO emission superposed on a
multi-band optical image created by the Hubble Heritage Team.  Here,
the quite different optical morphologies of the regions where CO
emission is found are evident more dramatically. This
composite image was created from different WFPC2 pointings in the
F336W (U), F502N ([O III]), F555W (V), F656N (H$\alpha$), F702N (R),
and F814W (I) filters.  A detailed discussion of these data is
published by MacKenty et al.\ (2000).

Both Beck, Turner \& Kovo (2000) and MacKenty et al.\ (2000) present
high resolution VLA radio continuum images.  There is 6 and 20 cm
emission associated with the central and SE star forming complexes,
but nothing is detected at the location of the NW CO emission (Turner
2000, private comm.). This lack of radio continuum emission
indicates that indeed very little star formation is associated with
the NW molecular region (C1--C8) and that the presence of only very
weak optical emission lines is not due to possible extinction.

\placefigure{fig7_rev}

\placefigure{fig8_rev}

\section{Discussion}

\subsection{The CO-H$_2$ Conversion Factor} 
\label{xfact}

To determine the conversion factor with a reasonable degree of
confidence, a discrete unit of emission, such as a GMC, must be
distinguishable in the data, and there must be a method to determine
the total mass associated with this feature which is independent of CO
intensity. GMCs are often assumed to be gravitationally bound and in
virial equilibrium, allowing the use of the virial theorem to
determine the total mass.  As discussed above, our spatial resolution
will not resolve GMC--sized objects at the distance of NGC~4214.  For
the sake of comparison, we decided to derive virial masses for our
clumps anyway -- in the discussion that follows we will keep in mind
that at least some of them are likely {\em not} gravitationally bound
objects. In the cases where the complex is not spatially resolved we
took the beamsize ($D\sim90$\,pc) as an upper limit for the actual
diameter -- then, the virial masses are probably upper limits (but see
the discussion below). If the virial mass could be trusted, dividing
it by the molecular mass derived using a Galactic CO-H$_2$ conversion
factor would yield the actual conversion factor for each feature, in
units of the Galactic value. The Galactic conversion factor is taken
to be $2.3
\times10^{20}$\,cm$^{-2}$\,(K\,km\,s$^{-1}$)$^{-1}$ (Strong et al.\
1988), with an additional explicit correction for helium (see caption
in Table 4). If the recalibrated conversion factor of $1.6
\times10^{20}$\,cm$^{-2}$\,(K\,km\,s$^{-1}$)$^{-1}$ (Hunter et al.\
1997) or the new value advocated by Dame et al.\ 2001 ($1.8
\times10^{20}$\,cm$^{-2}$\,(K\,km\,s$^{-1}$)$^{-1}$)  is used, all 
CO based masses are lower by $\sim 30$\%.  To facilitate comparison to
previous work, the numbers we give refer to the 'standard' value of
Strong et al.

The average conversion factor we find for the few resolved features is
compatible with the standard Galactic conversion factor (Tab.~4) given
all the uncertainties in deriving these numbers. The mean conversion
factor is $\sim 1.2 \pm 0.7 \times X_{\rm MW}$, whether or not only
the resolved clouds or all clouds are used for the estimate.  This is
a lower value than the conversion factor estimated for other dwarf
galaxies with similar metallicities in the literature ($2.7\ X_{\rm
MW}$ for the dIrr IC10, $< 2.2\ X_{\rm MW}$ for the dIrr NGC~6822,
both in the Local Group, Wilson 1995, and $6.6\ X_{\rm MW}$ in the
post-starburst dwarf galaxy NGC~1569, Taylor et al. 1999).

This result is unexpected at first glance since one would expect the
X--factor to be higher in this low metallicity galaxy (as suggested by
other studies of low--metallicity dwarfs, see above). We can only
speculate here why the results in NGC\,4214 are different. We should
like to note that the virial masses are strongly affected and possibly
even dominated by the resolution (both spatially and kinematically)
employed in the observations. E.g., if we use our naturally weighted,
5\,\kms\ channel width data (i.e., if we do not resolve the
substructure) we derive a virial mass for the NW complex which is
several times higher than the mass derived using $X_{MW}$. This is due
to fact that the dispersion in the NW region is rather large
($\Delta$v$\sim20$\,\kms) resulting in a much higher X--factor
(because of the $\Delta v^2$ dependence of the virial mass) -- in
other words our result would then be consistent with the values found
in other low--metallicity galaxies!  It should be also stressed here
that deriving cloud properties (and hence virial masses) is not at all
an objective process: e.g., different radii have been used by
different authors (e.g., diameter defined by the half--intensity line;
or by the contour containing 90\% of the flux) and the cloud sizes are
not always de--convolved for the beamsize. Dividing clumps into
sub--clumps is also always subject to to some personal bias.

It should be also noted at this point that convincing arguments have
been presented (e.g.\ Bolatto et al.\ 1999, Madden et al.\ 1997) for a
cloud structure in dwarf galaxies that significantly differs from what
is found in metal-rich giant spirals.  In this scenario, small CO
cores are embedded in huge molecular envelopes which are devoid of
CO. The molecular hydrogen in the envelopes coexists with \ion{C}{1} and
\ion{C}{2}. E.g., Madden et al.\ argue that in the dwarf irregular IC\,10
the total molecular mass may be up to 100 times the mass contained in
the CO cores. In this case our CO observations would only trace the
cores of the clouds and the derived `local' X--factor would be only
valid for the central core of a molecular cloud, where it might indeed
be close to the Galactic value, since the conditions in the core
region are expected to be similar to those in a Galactic GMC.

Unfortunately, based on the data presented here all this remains
subject to speculation. In the case of NGC\,4214 the matter might be
investigated further by a comprehensive study of the CO, \ion{C}{1},
\ion{C}{2}, radio continuum and FIR emission, along the lines of the
study Madden et al.\ performed for the (much closer) galaxy IC\,10.

Based on our data, we will adopt $X_{MW}$ as the conversion factor for
NGC~4214 in the following, primarily for convenience. We keep in mind
that this number is unfortunately only poorly determined and that our
molecular gas mass estimated might be off by a factor of 2 -- 3.

\subsection{Star Formation efficiency}

Given the fact that the SE molecular cloud seems not yet to be
disrupted by the ongoing burst of star formation (see the discussion
in Sec.~\ref{effects}), we can try to calculate the star formation
efficiency for that region. From Table 3 of MacKenty et al.\ (2000) we
derive a SFR for this region of 0.066\,M$_{\odot}$\,year$^{-1}$ (their
region II, corrected for `medium internal extinction'). For
morphological reasons (e.g.\ the compact and filled appearance of the
\ion{H}{2}\ regions, the lack of large cavities and the positional
coincidence of exciting star clusters and H$\alpha$ emitting knots),
star formation presumably only recently started here -- we adopted an
age of $\sim$3 Million years (MacKenty et al.\ 2000). This results in
a total stellar mass of $\sim 2.0\times10^5$\,M$_{\odot}$ produced to
date (assuming a constant SF rate). From our Table 4, we derive a
molecular mass of $24\times10^5$\,M$_{\odot}$ (summing up regions
C10--C14 and adopting a Galactic X--factor, see the discussion in
Sec.~\ref{xfact}). The star formation efficiency for the SE region is
thus around 8$\%$, somewhat higher than what is usually adopted (see
also, e.g., Wilson \& Matthews 1995 and the discussion in Taylor et
al.\ 1999).

\subsection{Star formation Threshold}

Studies of star forming dwarf galaxies suggest that the gas column
density plays an important role in regulating star formation (Skillman
et al.\ 1987, Taylor et al.\ 1994, Walter et al. 1997, van Zee,
Skillman \& Salzer 1998, Walter \& Brinks 1999).  A threshold column
density for the onset of star formation is thought to exist, at $\sim
10^{21}$ \cm2, above which massive star formation can proceed and
below which it is suppressed.  The physical reason for this threshold
may be related to the density needed for the shielding of molecular
gas from dissociating radiation (e.g. Skillman 1987), or perhaps to
disk instabilities (e.g.\ Kennicutt 1989).  For this kind of analysis,
generally some measure of the spatial extent of the star formation
activity (such as an H$\alpha$ image) is compared to the azimuthally
averaged gas surface density (usually based on \ion{H}{1} observations), or to
the contours of an \ion{H}{1} column density map.  Because CO is difficult to
detect in most dwarf galaxies, the previous studies have tended to
ignore the presence of molecular gas entirely, or else to attempt to
make a statistical correction by assuming a global H$_2$ to \ion{H}{1} ratio
(Hunter \& Plummer 1996).  However, we have high resolution CO and \ion{H}{1}
data on NGC 4214 covering the star forming regions of the galaxy, so
we can treat {\it both} the atomic and the molecular gas in the
threshold analysis.

We created a total gas column density map by combining the \ion{H}{1}
and CO moment maps, and using a Galactic CO-H$_2$ conversion factor
(see Sec.~\ref{xfact}).  Fig.~9 shows the contours for a column
density of 1, 2 and 4 $\times 10^{21}$ \cm2 (corrected for an
inclination of 30\deg; McIntyre 1997) for the total gas map and the
\ion{H}{1} map alone, superposed on the H$\alpha$ image from Fig.~7.
There is hardly any change in the lowest contour, indicating that at
densities of $10^{21}$ \cm2 the gas is primarily in atomic form.  The
higher contours, however, show marked changes from \ion{H}{1} alone to
total gas density, particularly at the location of the SE CO complex.
In the total gas map, we see a peak in gas density corresponding to
the group of \ion{H}{2} regions, whereas if we consider the \ion{H}{1}
alone, the H$\alpha$ emission sits in a local minimum of density,
adjacent to two local peaks.  The peak gas density in the total gas
map at this complex is 6.6\,$\times~10^{21}$ \cm2.  In the central
complex the addition of the molecular component also makes a
difference, connecting two \ion{H}{1} peaks that straddle the center
of the star forming complex.

\placefigure{fig9_rev}

From Fig.~9 it is apparent that there is only very faint H$\alpha$
emission associated with the CO emission of the NW complex, suggesting
very little star formation.  This lack of activity is confirmed by the
non-detection of radio continuum emission at the position of the NW
complex. The peak (\ion{H}{1} plus H$_2$) column density here is 95\% that of
the SE clump, where we see evidence for vigorous star formation.  The
column density in the central, very actively star forming region is
considerably lower, but this could be explained by the influence of
the massive stars upon the ISM in a more evolved star formation region
(see next section).  A plausible explanation for the lack of star
formation in the NW complex is suggested by Fig.~6, the comparison
of the CO and \ion{H}{1} velocities/linewidths.  Here we see that the
linewidth in CO in the NW complex is comparable to the \ion{H}{1} linewidth,
while for the other two complexes CO is much narrower.  Perhaps the
{\it surface} density along the line of sight to the NW complex is
high, but the {\it volume} density is not.  The large linewidth and
identifiable substructure in the NW complex suggests a distribution of
clouds at different velocities along the line of sight, rather than
one dense cloud.  It may also be that star formation has simply not
started yet in the NW complex in which case another episode of star
formation might begin in the near future.

\subsection{The Effects of the Star Formation Burst on the ISM}
\label{effects}
Massive stars, the most prominent constituents of star forming
regions, input energy into the ISM through mechanisms such as ionizing
photons, stellar winds and supernovae, and it is inevitable that they
will disrupt the local ISM.  In some dwarf galaxies, such as IC~2574,
swept out holes in the \ion{H}{1} accumulate large quantities of gas
on their rims, providing material for a second generation of star
formation (Walter et al.\ 1998, Stewart \& Walter 2000).  In NGC~4214
there is no evidence for this.  There are certainly a number of holes
in the \ion{H}{1} (Fig.~4) but the star forming regions do not appear
along their rims (as is apparent in the supergiant shell in IC~2574).
There are several small expanding shells of ionized gas associated
with the central star forming region (Martin 1998) and these might
later develop \ion{H}{1} shells like those seen in IC~2574.

On scales of a few hundred pc, the ISM shows indications of the
effects of the massive stars.  The \ion{H}{2}\ regions in the central
star forming complex show a shell morphology (Figs.~8 and~9) which
suggests that these shells have broken through the dense material
surrounding the massive stars.  The SE \ion{H}{2}\ regions still show
compact, knotlike morphologies, suggesting that a breakout has not yet
occurred (MacKenty et al.\ 2000).  If the starburst of the central
region preceded the SE one, then the massive stars in the center have
been at work dissociating and disrupting the molecular gas for a
longer time than in the SE.  Thus the CO intensity associated with the
shell \ion{H}{2}\ regions is lower and the emission is more diffuse
than $\sim 200$\,pc to the west, where the highest CO intensity in the
central complex is found, away from the immediate vicinity of any
\ion{H}{2}\ region.  The brightest CO emission in the galaxy, associated
with the \ion{H}{2}\ regions in the SE complex, may represent a
pre-disruption stage where the energy input into the ISM from the
young stars has not had enough time to affect a large fraction of the
molecular gas or the general morphology of the complex and is thus not
visible in our 90\,pc resolution observations.  In summary, NGC~4214
seems to show molecular gas and star formation in 3 distinct stages:
high--dispersion molecular gas with hardly any associated star
formation in the NW, concentrated molecular gas with very recent star
formation in the SE and diffuse molecular gas with an older massive
star formation burst in the center.

\section{Summary and Conclusions}

We have mapped the dwarf starburst galaxy NGC~4214 in the $^{12}$CO
(1--0) transition at the OVRO millimeter array with a linear
resolution of $\sim 90$\,pc. This first high-resolution study of the
molecular component in this galaxy shows the presence of three
molecular complexes which are in very different and distinct
evolutionary stages. Both from the properties of the molecular gas and
from HST and ground--based optical studies of the \ion{H}{2}\ regions and
star clusters, the central region is the most evolved: Here, the
molecular complex is highly disrupted, the impact of the massive star
formation is largest, and a shell structure is in the process of
forming. The star formation episode in the SE complex is more recent;
thus, the dense ISM adjacent to the newly formed massive stars is far
less disrupted, providing an ideal environment for ongoing, maybe
self-propagating star formation proceeding further into the
complex. Somewhat surprisingly, massive star formation has not started
yet in the NW complex, possibly because the threshold volume density
has not been reached in this region.

A comparison with the \ion{H}{1} reveals that no clear correlation
between the \ion{H}{1} and CO surface densities exists. Although the
NE and SW regions have almost the same total (\ion{H}{1} and H$_2$)
gas column density, violent star formation is only taking place in the
SW region.  This nicely illustrates the irregular nature of these
dwarf systems. The \ion{H}{1} and H$\alpha$ velocities are in good
agreement with the CO velocities measured in the three molecular
complexes. Assuming a Galactic conversion factor, the star formation
efficiency for the SE complex is $\sim 8$\%, similar to the one found
for spiral galaxies. This may be taken as an indication that the star
formation process itself in dwarf galaxies proceeds in a similar way
to that in spirals.

We were able to subdivide the molecular complexes further into 14
sub--complexes. Most of the clumps can not be resolved with our
synthesized beam -- even the resolved objects are probably not
GMCs. Deriving virial masses from these objects is not trivial and is
affected by personal bias and the employed velocity and spatial
resolution. Our clump analysis is compatible with a Galactic
conversion factor for NGC\,4214; this is substantially lower than the
value found in other dwarf galaxies with similar metallicities so
far. However, given the uncertainties in deriving this value we can not
rule out the validity of the metallicity--X relation based on our
data.  We derive a total molecular mass for NGC\,4214 of
5.1$\times10^6$\,M$_{\sun}$.

In summary, our OVRO observations of the molecular gas in the
metal--poor dwarf galaxy NGC\,4214 reveal an unexpected structural
wealth -- they serve as a good excuse to map similar low--metallicity
dwarf galaxies with interferometers in the future: even though the
emission is faint when compared to spiral galaxies, high--resolution
observations of the molecular gas can provide important insights to
the irregular nature of dwarf galaxies.

\acknowledgements

We thank C. Martin for providing us with her H$\alpha$ image of
NGC~4214.  FW acknowledges NSF grant AST96--13717. VJM was supported
in part by a Smithsonian Predoctoral Fellowship and an Australian
Postgraduate Research Award. Support of this work was also provided by
a grant from the K.T. and E.L. Norris Foundation. Research with the
Owens Valley Radio Telescope, operated by Caltech, is supported by NSF
grant AST96--13717.  The Five College Radio Astronomy Observatory is
operated with the permission of the Metropolitan District Commission,
Commonwealth of Massachusetts, and with the support of the NSF under
grant AST97--25951. The National Radio Astronomy Observatory (NRAO) is
operated by Associated Universities, Inc., under cooperative agreement
with the National Science Foundation.  The home of the Hubble Heritage
Project is Space Telescope Science Institute which is run by the
Association of Universities for Research in Astronomy for NASA. This
research has made use of the NASA/IPAC Extragalactic Database (NED)
which is operated by the Jet Propulsion Laboratory, Caltech, under
contract with the National Aeronautics and Space Administration
(NASA). The authors also acknowledge the use of NASA's Astrophysical
Data System Abstract Service (ADS), and NASA's SkyView.

\vfill
\eject
\begin{center}
{\bf Figure Captions}
\end{center}

\figcaption[walter.fig1_rev.ps] {Integrated intensity maps of CO (1--0) emission
 in NGC~4214 (grayscale and contours).  The left panel shows the
 naturally weighted map (6\arcsec.5~$\times$~5.7\arcsec\
 (131$\times$117\,pc) resolution) and the right panel shows the
 robustly weighted map (4\arcsec.8~$\times$~4.2\arcsec\
 (98$\times$86\,pc) resolution).  The contours show 10, 20, 40 and
 80\% of the peak intensity (natural: 5.0 Jy\,beam$^{-1}$\,\kms;
 robust: 3.9 Jy\,beam$^{-1}$\,\kms).
\label{fig1_rev}}

\figcaption[walter.fig2_rev.ps] {Position--velocity diagram indicating the subdi
visions
of the CO emission in the NW complex (ellipses). The numbers refer to
the clumps as listed in Table~4. The position velocity cut is centered
on RA=12:15:36.5, DEC=36:20:02 (J2000), the position angle is
270$^\circ$ (east--west).
\label{fig2_rev}}

\figcaption[walter.fig3_rev.ps] {The diameter--line width relationship for GMCs 
in Local Group galaxies.  The clouds in different galaxies are plotted
as different symbols, and the line shows a best fit relationship to the M33
GMCs derived by Wilson \& Scoville (1990).  Upper limits are given for
the spatially unresolved objects in NGC~4214.
\label{fig3_rev}}

\figcaption[walter.fig4_rev.ps] {{\it Left:} The integrated intensity of \ion{H}
{1} in NGC~4214.
The center box indicates the region shown in the right panel. {\it
Right:} The naturally weighted CO contours from Fig.~1 superposed on
the \ion{H}{1} integrated intensity map. The line indicates the
orientation of the position--velocity cut presented in Fig.~6 (right).
\label{fig4_rev}}
 
\figcaption[walter.fig5_rev.ps] {Cleaned channel maps of CO emission (contours, 
convolved to 9$''$ resolution) superposed on the same channels of
\ion{H}{1} emission (greyscale). The contours show 2.5, 5 and
7.5$\sigma$, where 1$\sigma$\,=\,40\,mJy\,beam$^{-1}$. Channel
seperation is 1.3\,\kms.
\label{fig5_rev}}

\figcaption[walter.fig6_rev.ps] {{\it Left:} Position velocity diagram of 
the \ion{H}{1} distribution (centre: RA=12:15:38.5, DEC=36:19:34
(J2000), position angle: 319$^\circ$). The center
box indicates the region shown in the right panel. {\it Right:}
\ion{H}{1} position velocity diagram with CO contours (white)
superposed (see the line in Fig.~4, left, for the orientation of this
cut).
\label{fig6_rev}}

\figcaption[walter.fig7_rev.ps] {{\it Left:} Image of H$\alpha$ + [NII] emission
 in
NGC~4214 with the contrast set to show the low surface brightness
features.  The box indicates the region shown in the right panel.  The
gray scale has been chosen to bring out faint H$\alpha$ filaments in
addittion to the bright \ion{H}{2}\ regions.  {\it Right:} The naturally
weighted CO contours from Fig.~1 superposed on the H$\alpha$ + [NII]
emission.  The large circles show the area covered in our CO
mosaic. The gray scale is set lower to show the \ion{H}{2}\ regions without
saturation. Only a very faint \ion{H}{2}\ region is visible at the position
of the NW CO complex, while strong H$\alpha$ emission is obvious at
the positions of the central and SE complexes.
\label{fig7_rev}}

\figcaption[walter.fig8_rev.ps] {The naturally weighted CO contours from Fig.~1 
superposed on a multi-wavelength F336W (U), F502N ([O III]), F555W
(V), F656N (H$\alpha$), F702N (R), F814W (I) image created by the
Hubble Heritage Team from WFPC2 data.
\label{fig8_rev}}

\figcaption[walter.fig9_rev.ps] {{\it Upper Left:} \ion{H}{1} column density cor
rected for 
inclination, with contour levels at 1 and 2$\times~10^{21}$ \cm2.
{\it Upper Right:} Total gas (\ion{H}{1} + H$_2$) column density
corrected for inclination, with contour levels of 1, 2, and 4
$\times~10^{21}$ \cm2. To faciliate the interpretation of
the contours in the upper two panels, greyscale plots are presented at
the bottom.
\label{fig9_rev}}

\clearpage

\begin{deluxetable}{ll}
\tablewidth{220pt}
%\vspace*{-2in}
\tablecaption{Properties NGC 4214  \label{tab2}}
\tablehead{ \colhead{Property} & \colhead{Value} }
\startdata
%Property & value \nl
R.A. (2000) & 12:15:39.1 \nl
Dec. (2000) & 36:19:39   \nl
V(sys)$_{LSR}$ & 300 \kms \nl
Distance$^a$ & 4.1 Mpc \nl
1$''$ & 20 pc \nl
M$_{HI}^b$ & 1.0~$\times~10^9$ M$\odot$ \nl
B$_T^c$, M$_B^c$ & 10.2, -18.8 \nl
12 + log(O/H)$^d$ & 8.22 \nl 
L$_{FIR}^e$ & 6.3~$\times~10^8$ $\odot$ \nl
Star formation rate$^f$ & $\sim$ 0.4 M$\odot$ yr$^{-1}$ \nl
\enddata
\tablecomments{$^a$Leitherer et al. (1996), $^b$Swaters (1999),
$^c$de Vaucouleurs et al. (1991), $^d$Kobulnicky \& Skillman (1996),
$^e$Thronson et al. (1988), $^f$MacKenty et al. (2000).
 }

\end{deluxetable}

\clearpage

\begin{deluxetable}{lcc}
\tablewidth{300pt}
\tablecaption{Summary of the OVRO observations  \label{tab3}}
\tablehead{}
\startdata
configuration & C & L \nl
baselines & $7-22\,\mbox{k}\lambda$ & $6 - 44\,\mbox{k}\lambda$ \nl 
          & $ (20-60 \,\mbox{m})$ & $ (15-115\,\mbox{m})$ \nl
\tableline
tracks    & 1999 Sep 27 & 1999 Nov 9  \nl
          & 1999 Sep 28 & 1999 Oct 10 \nl
          & 1999 Oct 2  & 2000 Mar 27 \nl
          & 2000 Jun 2  & 2000 Apr 11 \nl
          & 2000 Jun 3  & 2000 Apr 12 \nl
          & 2000 Jun 4  & 2000 Apr 29 \nl
          &             & 2000 May 6  \nl
\tableline
\tableline
                         & Setup 1  & Setup 2 \nl   
total bandwidths         &  124 MHz & 31 MHz   \nl
No. of channels          &  62      & 62   \nl
velocity resolution      &  5\,\kms & 1.3\,\kms  \nl
\tableline
                        & natural weighting   & robust weighting \nl
angular resolution      & $6.4''\times 5.7''$ & $4.8''\times 4.2''$\nl
linear resolution$^{a}$ & $131 \times 117$ pc & $98 \times 86$\nl
rms noise$^{b}$         & 40 mJy/beam         & 47 mJy/beam \nl
                        & 100 mK              & 210 mK \nl

\enddata
\tablecomments{\\$^a$ adopting a distance of 4.1\,Mpc.\\
$^b$ noise is given for a 1.3\,\kms\ resolution channel.}
\end{deluxetable}

\end{document}